\documentclass[a4paper]{jpconf}
\usepackage{graphicx}

\begin{document}
\title{The ArDM project: a Liquid Argon TPC for Dark Matter Detection\footnote{
Based on an invited talk given at the Third Symposium On Large TPCs For Low Energy Rare Event Detection, 11 - 12 December 2006, Paris (France).}}

\author{Marco Laffranchi, Andr\'e Rubbia, on behalf of ArDM collaboration}

\address{Institute for Particle Physics, ETH Zurich}

\ead{marco.laffranchi@cern.ch, andre.rubbia@cern.ch}

\begin{abstract}
WIMPs (Weakly Interacting Massive Particles) are considered the main candidates for Cold Dark Matter. The ArDM experiment aims at measuring signals directly induced by WIMPs in liquid argon. A 1-ton prototype is currently developed with the goal of demonstrating the feasibility and performance of a detector with such a large target mass. ArDM aims at acting as a liquid argon TPC and additionally measuring the scintillation light.
The principle of the experiment and the conceptual design of the detector are described.
\end{abstract}

\section{Introduction}

The existence of Dark Matter has been postulated in order to describe astronomical observations. This kind of matter interacts with ordinary matter only gravitationally and presumably weakly but not electromagnetically, so that it looks ''dark'' and manifests itself only indirectly by its impact on visible structures like galaxies, galaxy clusters or gravitational lensing. Common models for Dark Matter describe it as consisting of non-baryonic massive particles or WIMPs (Weakly Interacting Massive Particles) forming a cold thermal relic gas.\\
In the last years great effort has been put into direct observation of WIMPs, which would validate the Dark Matter model. Due to their weak coupling and their low kinetic energy, interactions of WIMPs are rare and elusive events. The recoil of a nucleus in a target with a recoil energy ranging between 10~keV and 100~keV can be imparted by a WIMP interaction. The total interaction rate depends on underlying models, e.\,g.~supersymmetric models, whose parameters are numerically unknown, and is therefore not yet predictable.\\

In 2004, the Argon Dark Matter experiment (ArDM) was initiated \cite{Proc}. The goal of this project is the development of a 1-ton argon detector with independent ionization and scintillation readout, demonstrating the feasibility of a ton-scale liquefied noble gas Dark Matter experiment. These are unique features compared to other nobles liquid projects \cite{XENON, WARP1, ZEPLIN, ZEPLIN2}, (see also \cite{Proc2}).

Liquid noble elements like argon or xenon are excellent targets for Dark Matter detection \cite{IC1,IC2}.
Their high density, together with the Time Projection Chamber (TPC) technique, characterizes the detector as a finely granulated, homogeneous calorimeter. The choice of natural argon instead of xenon for the initial ton-scale target is motivated by the fact that the event rate in argon is less sensitive to the energy threshold than in xenon, due to form factor effects. Argon is cheaper than xenon. A ton-scale argon detector is hence readily conceivable, safe and economically affordable and can act as a demonstration for large Dark Matter noble liquid detectors. Furthermore, the recoil spectra are different in xenon and argon, providing a crosscheck if a potential WIMP signal is measured.\\

Argon and xenon scintillate with a high photon yield, giving at the same time a prompt signal for the time resolution of the event. Furthermore, the scintillation light and ionization charge provide an indication of the kind of the interacting particle. 
One possibility for particle recognition is achieved by measuring the ratio of the scintillation light and the charge remained after recombination; both signals strongly depend on the ionization density induced by the interacting particle. Nuclear recoils induced by WIMPs and neutrons produce a high ionization density compared with $\gamma$ and electron interactions and can thus be distinguished. Additionally, the fast and slow component of the scintillation light of the liquid argon are distinguishable, and their relative strength also depends on the ionization density---facilitating further independent particle identification \cite{Boulay:2006mb}. A remaining background are fast neutrons, which can induce nuclear recoils indistinguishable from a WIMP interaction. Therefore, an external neutron moderator and  shielding need to be added and special care has to be taken concerning the radiopurity of the materials in the inner detector \cite{Backgr}. With a large fiducial volume multiple scattering of neutrons inside the target is more probable, which makes them distinguishable from a WIMP interaction. Nevertheless, it is important to reduce background radiation to a very low level. WIMP Direct Detection experiments therefore need to be located underground. Summarizing the four means for background rejection:

\begin{itemize}
\item{Light--charge ratio of a localized event}
\item{Fast and slow component of the scintillation light}
\item{Topology of the event and multiple scatter event rejection} 
\item{Using liquid argon as an active shield by reducing the fiducial volume} 
\end{itemize}

\section{Design of the ArDM detector}

\noindent

\begin{figure}
\begin{center}
\setlength{\unitlength}{0.9mm}
\begin{picture}(100,130)
\put(3,3){\includegraphics[width=60\unitlength]{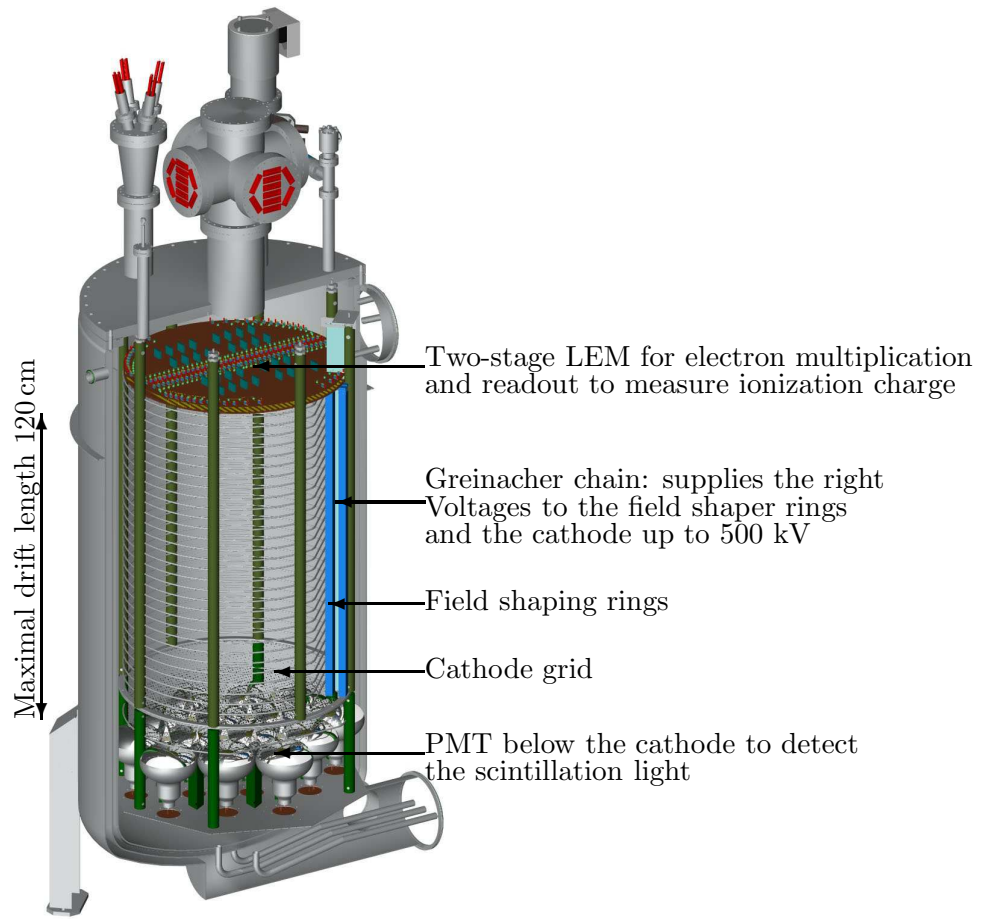}}
\thicklines

\put(60,85){\vector(-1,0){25}}
\put(60,88){\makebox(0,0)[tl]{Two-stage LEM for electron multiplication}}
\put(60,84){\makebox(0,0)[tl]{and readout to measure ionization charge}}

\put(60,65){\vector(-1,0){13}}
\put(60,70){\makebox(0,0)[tl]{Greinacher chain: supplies the right}}
\put(60,66){\makebox(0,0)[tl]{Voltages to the field shaper rings}}
\put(60,62){\makebox(0,0)[tl]{and the cathode up to 500~kV}}

\put(60,50){\vector(-1,0){14}}
\put(60,52){\makebox(0,0)[tl]{Field shaping rings}}

\put(60,40){\vector(-1,0){22}}
\put(60,42){\makebox(0,0)[tl]{Cathode grid}}

\put(60,28){\vector(-1,0){22}}
\put(60,31){\makebox(0,0)[tl]{PMT below the cathode to detect}}
\put(60,27){\makebox(0,0)[tl]{the scintillation light}}

\put(4,33){\vector(0,1){45}}
\put(4,78){\vector(0,-1){45}}
\put(0,33){\rotatebox{90}{Maximal drift length $120$\,cm}}

\end{picture}
\caption{\label{fig} Setup of the ArDM detector.}
\end{center}
\end{figure}

Fig.~\ref{fig} illustrates the conceptual design of the ArDM 1-ton prototype. An ionizing particle induces prompt scintillation light and free ionization electrons in the liquid argon volume. The electrons are drifted upwards to the surface by an electric field. The maximal drift length is about 120~cm. A Greinacher (Cockcroft--Walton) circuit provides the correct potential to the field shaper rings and the cathode grid in order to reach an electric field strength up to 4~kV/cm inside the target with a good homogeneity. Just below the liquid argon surface a metallic grid is located. It increases the electric field at the surface in order to extract the electrons from the liquid to the gas phase. Once in argon gas, the electrons are multiplied by a two stage Large Electrons Multiplier (LEM). The LEM consists of a printed board with metallization on both sides and equally spaced holes. If a sufficient potential difference is applied, the electrons are drawn into the holes where the high electric field induces an avalanche multiplication. The metallizations of the second stage are striped and act as a position sensor of the TPC. Each strip is independently read by a low noise charge sensitive preamplifier with a typical equivalent noise of $\simeq$ 1000 electrons.\\

For the readout of the scintillation light, an array of 14 photomultiplier tubes is foreseen, which is located below the cathode grid. The primary UV scintillation light has a wavelength of $\sim$128~nm and needs to be shifted to visible light in order to match the sensitivity range of the photomultiplier tubes. A reflector coated with TPB (Tetra-Phenyl-Butadiene) wavelength shifter surrounds the fiducial volume and increases the light collection efficiency. The PMT glass have also a specially optimized coating of TPB with good transparency to the visible light and at the same time an high wavelength shifting efficiency.\\

The liquid argon must be very pure in order to prevent electron loss due to dissolved electronegative impurities. A system of pump and filter purifies the liquid to the ppt-level purity. The cryostat, the recirculation system and the feedthroughs are designed to minimize the heat input. Furthermore, the liquid argon vessel is in thermal contact with a thermal bath in order to avoid bubbles that could trigger discharges at the high voltage stages. This system has been designed and built in collaboration with Bieri engineering GmbH. The temperature, pressure, precision levelmeter and a liquid argon purity monitor are part of the slow control.\\

A powerful distinction between lowly and highly ionizing  particles is a mandatory condition for the rejection of predominant backgrounds like the internal $^{39}$Ar signal. The $^{39}$Ar isotope is present in natural argon liquefied from the atmosphere \cite{Loosli, Ar39}, and produces a background rate due to beta decay of approximately 1 kHz in one ton of liquid argon. The alternative possibility of using $^{39}$Ar-depleted argon extracted from underground natural gas is currently also studied. \\

\section{Outlook}

Over the last years, a lot of efforts have been undertaken to find out more about the nature of Dark Matter as well as the particle which it might consist of. It is believed that liquefied noble gas detectors provide one approach to identify the WIMP. The ArDM project aims at the application of the liquid argon TPC for Direct Detection of nuclear recoils induced by WIMPs. The first goal of the project is the construction of a 1-ton prototype in order to demonstrate the feasibility of the experiment. The three keypoints for a successful operation are:

\begin{itemize}
\item{LEM-based charge readout in gaseous argon}
\item{Generation of a very high drift field}
\item{Efficient readout of the argon scintillation light} 
\end{itemize}

The 1-ton prototype is currently under construction at CERN. Its operation involves the design, acquisition and run of a cryogenic system, a liquid argon purification system, a high-voltage system, charge amplification and readout, and light readout. 
The first milestone to be achieved during 2007 is a proof of principle and stability studies, as well as assessment of the detector performance regarding $\gamma$-ray and beta electron background rejection versus nuclear recoils.\\

 After the successful construction, testing and operation of the prototype, an underground operation is planned in order to minimize background radiation induced by cosmic rays. At this stage, the design of a moderator and shielding against neutrons from detector components, surrounding facilities and muon-induced neutrons will be addressed.\\
 
With a recoil energy threshold of 30~keV, a WIMP-nucleon cross section of $10^{-6}$~pb would yield approximately 100 events per day per ton. The sensitivity of the ArDM 1-ton prototype would therefore access the WIMP-nucleon cross section region of $10^{-6}$~pb. By improving the background rejection power and further limiting the background sources, a sensitivity of $10^{-8}$~pb would become reachable. Scaling linearly with mass, a 10-ton detector could reach a ten times smaller cross section. Due to the scalable technologies used and the low cost of argon, an enlargement of the volume is a realistic prospect. 


\section{Acknowledgements}

The help of all ArDM colleagues from ETH Z\"urich, Granada University, CIEMAT, Soltan Institute Warszawa, University of Sheffield and University of Z\"urich is greatly acknowledged. Informal contributions from P. Picchi (LNF) are also greatly recognized.

\end{document}